\documentclass[iop,apjl]{emulateapj}
\begin{document}

\newcommand{\kms}{\ensuremath{\mathrm{km}\,\mathrm{s}^{-1}}}
\newcommand{\etal}{et al.}
\newcommand{\LCDM}{$\Lambda$CDM}
\newcommand{\ML}{\ensuremath{\Upsilon_{\star}}}


\title{The Baryon Content of Cosmic Structures}

\author{Stacy S. McGaugh\altaffilmark{1}, James M. Schombert\altaffilmark{2}, W.J.G. de Blok\altaffilmark{3},
and Matthew J. Zagursky\altaffilmark{4}}

\altaffiltext{1}{Department of Astronomy, University of Maryland, College Park, MD 20742-2421, USA; 
ssm@astro.umd.edu}
\altaffiltext{2}{Department of Physics, 1274 University of Oregon, Eugene, OR 97403-1274, USA;
jschombe@uoregon.edu}
\altaffiltext{3}{Department of Astronomy, University of Cape Town, Private Bag X3, 
Rondebosch 7701, Republic of South Africa;  edeblok@ciricnus.ast.uct.ac.za}
\altaffiltext{4}{Institute for Astronomy, University of Hawaii, 2680 Woodlawn Avenue, Honolulu, HI 96826, USA;
mzagursk@ifa.hawaii.edu}

\begin{abstract}
We make an inventory of the baryonic and gravitating
mass in structures ranging from the smallest galaxies
to rich clusters of galaxies.  We find that the
fraction of baryons converted to stars reaches a
maximum between ${M}_{500} = 10^{12}$ and $10^{13}\;\mathrm{M}_{\sun}$, 
suggesting that star formation is most efficient in bright galaxies in groups.
The fraction of baryons detected in all forms deviates monotonically 
from the cosmic baryon fraction as a function of mass. 
On the largest scales of clusters, most 
of the expected baryons are detected, while in the
smallest dwarf galaxies, fewer than 1\% are detected. 
Where these missing baryons reside is unclear.
\end{abstract}

\keywords{cosmological parameters --- dark matter --- galaxies: clusters: general --- galaxies: dwarf ---
galaxies: irregular --- galaxies: spiral}

\section{Introduction}

The early universe was a highly uniform and homogeneous mix of dark and baryonic matter
with baryon fraction ${f_b = 0.17 \pm 0.01}$ \citep{WMAP5} .  
If this primordial mix persists as individual gravitationally bound structures
emerge during the course of cosmic evolution, then the baryonic mass of any given object
would be $M_b = f_b M_{tot}$.  Indeed, combining this with the baryon density constraint from big
bang nucleosynthesis \citep{BBN} provides one important argument that the density parameter 
is less than unity \citep{White93,OS}.  Here we reverse the logic and ask what fraction of the
expected baryons are actually detected.

We present an inventory of the baryonic and gravitating masses of cosmic structures spanning
a dozen decades in detected baryonic mass.  The mass of baryons known in each system 
correlates well with the total mass, but not as a simple proportion.  The implication is that
most of the baryons associated with individual dark matter halos are now missing.
While some are in the intergalactic medium \citep{OVI}, a complete accounting
of where the baryons now reside, and how they relate to their parent structures, remains wanting.

\section{Mass Inventory} 

We divide our inventory into two broad categories of gravitationally bound systems:
those supported by rotation, and those supported by random motions.
This distinction is important to the way in which we infer the gravitating mass,
as discussed for each type of system below.  The detected baryonic mass is the sum of observed
stellar and gaseous components: $M_b = M_{*} + M_g$.  In general, the gas mass is more
precisely known as the physics of the emission mechanism is better understood, 
while the stellar mass requires an estimate of the mass-to-light ratio of a 
composite stellar population in order to convert observed luminosity to mass.
Over the many decades in mass considered here, a factor of $\sim 2$ uncertainty in the
the stellar mass-to-light ratio is a minor concern.

\begin{figure}
\epsscale{1.0}
\plotone{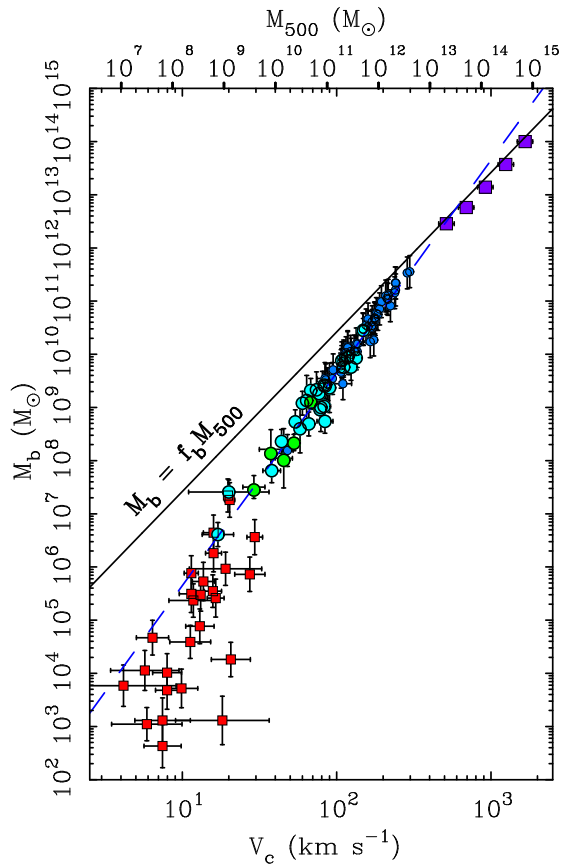}
\caption{{The relation between baryonic mass and rotation velocity.}
The sum of detected baryonic mass is plotted against the circular
velocity of gravitationally bound extragalactic systems.  Round symbols 
represent rotationally supported disks while square symbols represent
pressure supported systems.  Larger symbols correspond to systems
whose baryonic mass is dominated by gas and smaller symbols those
dominated by stars.  Dark blue circles are 
for star dominated spirals \citep{mcg05}.  Light blue \citep{stark} and green \citep{trach} 
circles represent recent work on gas dominated disks.  Red squares represent
Local Group dwarf satellites \citep{walker}.  
Purple squares represent the mean of many galaxy clusters \citep{giodini}.
These pressure supported systems fall close to, but systematically below the
Baryonic Tully-Fisher relation defined by the disks (dashed line).
The mass enclosed by an over-density 500 times the critical density is shown
by the upper abscissa assuming 
${M_{500} = (1.5 \times 10^5 \;\textrm{km}^{-3}\,\textrm{s}^{3}\,\textrm{M}_{\odot}^{-1}) V_c^3}$ 
(see discussion in text).  If structures possessed baryons in the cosmic fraction
(${f_b = 0.17}$), they would fall along the solid line.
\label{MbV}}
\end{figure}

\placefigure{MbV}

\subsection{Rotationally Supported Systems}  
 
 One of the most important lines of evidence for
dark matter is the observation that the rotation curves of spiral galaxies tend to become 
asymptotically flat \citep{vera,bosma} at large radii where the baryonic mass alone should
result in a Keplerian decline.  This outermost circular velocity, $V_c$, we take to 
be representative of the gravitating mass.   
Among rotating systems, we distinguish between those dominated by
stellar mass (typically early type spirals) and those dominated by gas
mass (typically irregular and late type spirals).

\subsubsection{Star Dominated Spiral Galaxies} 

Bright spiral galaxies have the majority of their detected
baryonic mass in the form of stars.  We utilize the rotation velocities
and mass estimates of \citet{mcg05}, selecting only those galaxies with
$M_* > M_g$ for $Q=1$.
The results are in good agreement with independent work \citep{THINGS}.
The sum of baryonic mass is completed by adding the cold gas 
mass, corrected by a factor of 1.4 to account for the presence of helium and heavier elements.
Dust and hot gas do not contribute significantly to the total \citep{bregman,andbreg}.

\subsubsection{Gas Dominated Galaxies} 

Many dim, late type disk galaxies have more of their baryonic
mass in the form of atomic gas than in stars: $M_* < M_g$.  
These are particularly interesting for our purposes here
because their baryonic mass is insensitive to the choice of stellar mass estimator.  For these
galaxies, $M_b$ follows almost directly from the observed 21 cm flux and the physics of the
hydrogen hyperfine transition.  Two recent independent studies \citep{stark,trach} provide 
for the first time a large combined sample of such galaxies.

Rotationally supported disk galaxies are known to obey a relation between baryonic mass and 
rotation velocity known as the Baryonic Tully-Fisher relation \citep{mcg05}.  This is shown in 
Fig.~\ref{MbV}, giving emphasis to the gas dominated galaxies (larger circles) which extend
the relation beyond that previously known for bright spirals.  They also provide a physics-based
calibration of the relation \citep{stark,trach}.  This is shown as the dashed line in Fig.~\ref{MbV}.  
The intrinsic relation appears to be very tight \citep{verhTF}, with observational uncertainty
(due mostly to the uncertainty in the distances to the individual galaxies) accounting for 
essentially all of the scatter.

\placetable{broken}
\begin{deluxetable}{lccc}
\tablewidth{0pt}	
\tablecaption{Broken Power Law Fit}
\tablehead{
\colhead{Scale} & \colhead{Range in $V$} & \colhead{$x$} & \colhead{$A$}
}
\startdata
Dwarf & $< 20$ & 8.2 & $-$3.88 \\
Disk & 20 --- 350 & 4.0 & $\phn{}$1.65 \\
Cluster & $> 350$ & 3.2 & $\phn{}$3.69
\enddata
\tablecomments{$\log M_b = x \log V_c + A$}
\label{broken}
\end{deluxetable}

The Baryonic Tully-Fisher relation for rotating disks shown in Fig.~\ref{MbV} extends over
five decades in baryonic mass, a considerable improvement 
over the two decades typically considered.  Before
considering how the baryonic mass relates to the total mass, we first investigate whether
this relation can be extended still further.  Rotationally supported systems cover
the range $20 < V_c < 300\;\textrm{km}\,\textrm{s}^{-1}$, but there are
both smaller and larger pressure supported systems.

\placetable{bindat}
\begin{deluxetable*}{lccr@{.}lcr@{.}lcr@{.}lcr@{.}lr@{.}lr@{.}lr@{.}lc}
\tabletypesize{\footnotesize}
\tablewidth{0pt}	
\tablecaption{Binned Data}	
\tablehead{
\colhead{System} &
  \colhead{$<$$V_c$$>$}  & \colhead{$\sigma_V$} &
   \multicolumn{2}{c}{$<$$M_b$$>$}  & \colhead{$\sigma_{b}$} & 
   \multicolumn{2}{c}{$<$$M_*$$>$}  & \colhead{$\sigma_{*}$} 
& \multicolumn{2}{c}{$<$$M_{500}$$>$} & \colhead{$\sigma_{M}$} 
& \multicolumn{2}{c}{$f_d$} & \multicolumn{2}{c}{$\sigma_f$}
& \multicolumn{2}{c}{$f_*$} & \multicolumn{2}{c}{$\sigma_*$}  & \colhead{Ref.} 
}
       \startdata	
Cluster &3.22   &0.05   &14&00  &0.05   &13&13  &0.11   &14&85  &0.15   &0&83   &0&04   &0&11   &0&01   & 1 \\
Cluster &3.10   &0.05   &13&57  &0.07   &12&80  &0.10   &14&48  &0.15   &0&73   &0&05   &0&12   &0&01   & 1 \\
Cluster &2.96   &0.05   &13&14  &0.04   &12&64  &0.11   &14&08  &0.15   &0&68   &0&03   &0&21   &0&02   & 1 \\
Cluster &2.84   &0.05   &12&76  &0.04   &12&36  &0.04   &13&71  &0.15   &0&67   &0&03   &0&26   &0&01   & 1 \\
Cluster &2.71   &0.05   &12&46  &0.21   &12&12  &0.08   &13&32  &0.15   &0&80   &0&16   &0&37   &0&03   & 1 \\
Spiral  &2.40   &0.05   &11&32  &0.17   &11&26  &0.14   &12&51  &0.15   &0&38   &0&07   &0&34   &0&05   & 2 \\
Spiral  &2.32   &0.03   &10&99  &0.09   &10&94  &0.08   &12&26  &0.08   &0&32   &0&03   &0&28   &0&02   & 2 \\
Spiral  &2.23   &0.02   &10&63  &0.09   &10&59  &0.10   &12&01  &0.07   &0&25   &0&02   &0&22   &0&02   & 2 \\
Spiral  &2.15   &0.06   &10&26  &0.14   &10&15  &0.20   &11&76  &0.18   &0&19   &0&03   &0&14   &0&03   & 2 \\
Spiral  &2.08   &0.04   &10&00  &0.06   &9&85   &0.12   &11&56  &0.11   &0&16   &0&01   &0&11   &0&01   & 2 \\
Gas Disk &2.07   &0.04   &9&85   &0.09   &9&49   &0.09   &11&53  &0.11   &0&12   &0&01   &0&053  &0&005  & 3 \\
Spiral  &2.03   &0.02   &9&79   &0.09   &9&62   &0.07   &11&39  &0.07   &0&15   &0&01   &0&10   &0&01   & 2 \\
Spiral  &1.92   &0.10   &9&31   &0.46   &9&18   &0.45   &11&06  &0.31   &0&10   &0&05   &0&077  &0&035  & 2 \\
Gas Disk &1.88   &0.05   &9&21   &0.17   &8&61   &0.34   &10&95  &0.16   &0&11   &0&02   &0&027  &0&009  & 3 \\
Gas Disk &1.78   &0.08   &8&62   &0.12   &7&79   &0.37   &10&64  &0.23   &0&057  &0&007  &0&0083 &0&0031 & 3 \\
Gas Disk &1.65   &0.14   &8&24   &0.58   &7&12   &0.35   &10&26  &0.42   &0&056  &0&032  &0&0043 &0&0015 & 4  \\
Gas Disk &1.37   &0.18   &7&28   &0.44   &6&74   &0.57   &9&42   &0.55   &0&042  &0&019  &0&012  &0&007  & 3 \\ 
Dwarf   &1.29   &0.10   &6&67   &0.54   &6&67   &0.54   &9&19   &0.29   &0&018  &0&010  &0&018  &0&0010 &  5 \\
Dwarf   &1.16   &0.13   &5&60   &0.20   &5&60   &0.20   &8&80   &0.38   &0&0037 &0&0007 &0&0037 &0&0007  & 5 \\
Dwarf   &0.94   &0.20   &3&81   &0.69   &3&81   &0.69   &8&13   &0.60   &0&0003 &0&0002 &0&0003 &0&0002  & 5
	\enddata
\tablerefs{1.\ Clusters: \citet{giodini}; 2.\ Spirals: \citet{mcg05}; Gas dominated galaxies:
3.\ \citet{stark} and 4.\ \citet{trach}; 5.\ Local Group dwarfs: \citet{walker}.}
\tablecomments{Velocities and masses are logarithmic with units of km s$^{-1}$ and solar masses,
respectively.}
\label{bindat}
\end{deluxetable*}	

 \subsection{Pressure Supported Systems}  
 
\subsubsection{Elliptical Galaxies}

Observations of giant elliptical galaxies typically do not extend far enough radially to 
identify the equivalent of $V_c$ \citep{sauron}.
Gravitational lensing \citep{hank,slacs} provides important constraints on the total mass, 
but the conversion to the equivalent circular velocity is sensitive to the assumed model.  
It is also subject to degeneracy between stellar and dark mass.
Given these uncertainties, and that these data largely overlap with those for spirals,
we do not consider giant ellipticals further.  We do note that the estimates of \citet{hank}
and \citet{slacs} are consistent with our results to the same degree that they are consistent with
each other.

\subsubsection{Local Group Dwarfs} 

Recent years have witnessed an explosion in the discovery of small satellite
galaxies of both the Milky Way \citep{newdwarfs} and M31 \citep{newAND}.  
The proximity of these quasi-spherical satellite
galaxies makes it possible to identify isolated stellar 
systems much smaller than known elsewhere,
and to measure their internal kinematics with velocities from
individual stars \citep{W07,walker,SG}.  We assume isotropic internal orbits
to relate the observed line-of sight velocity dispersion $\sigma$ to the circular 
velocity: $V_c = \sqrt{3} \sigma$ \citep{boom}.  
The baryonic masses of these systems are dominated by stars.  
To convert light to mass, we adopt the mean stellar mass-to-light ratio estimated \citep{mario} 
from resolved stellar population studies:
$M_{*}/L_V = 1.3\;\textrm{M}_{\odot}/\textrm{L}_{\odot}$.  

\subsubsection{Clusters of Galaxies} 

Recent work \citep{giodini} on clusters provides a large, 
homogeneous set of data with stellar, gas, and
gravitating masses averaged over many individual clusters.  Most of the baryonic mass in
clusters is in hot, X-ray emitting gas.  This provides
a good measure of both the gas mass and the gravitating
mass.  Stellar masses are based on a scaling relation \citep{giodini}
derived from $K$-band data.  The result here is not sensitive to the precise relation as the
X-ray gas dominates the baryon budget.  

It is conventional to refer to structures by the density contrast
they represent with respect to the critical density of the universe.  The mass
enclosed within a radius encompassing the over-density $\Delta$ is
$M_{\Delta} = ({4 \pi}/{3}) \Delta \rho_{crit} R_{\Delta}^3$.
With the definition of critical density and circular velocity, 
$M_{\Delta} = (\Delta/2)^{-1/2} (G H_0)^{-1} \; V_{\Delta}^3$.
The cluster data \citep{giodini} are referenced to $\Delta = 500$, so for
$H_0 = 72\;\textrm{km}\,\textrm{s}^{-1}\,\textrm{Mpc}^{-1}$ \citep{HSTKP},
$M_{500} = (204552 \;\textrm{km}^{-3}\,\textrm{s}^{3}\,\textrm{M}_{\odot}^{-1}) V_{500}^3$.
To relate the circular velocity $V_c$ observed in spirals to $V_{500}$
at larger radii, we note that rotation curves are approximately flat, so we expect 
$V_c = f_V V_{500}$ with $f_V \approx 1$.  If we associate the observed $V_c$ with
the peak velocity of an NFW halos \citep{NFW}, then 
$1.0 \le f_V \le 1.3$ over the relevant range of masses.  This appears consistent with the 
Milky Way \citep{flynn,xue,mcgMW} if $f_V = 1.1$, which we adopt for specificity.  This is also
consistent with the independent estimate of \citet{LNK}.  Note that a 20\%
change in $f_V$ corresponds to shifting the cluster data by the width of a data point in 
Fig.~\ref{MbV}.

\placefigure{fdfsV}
\begin{figure*}
\epsscale{1.0}
\plotone{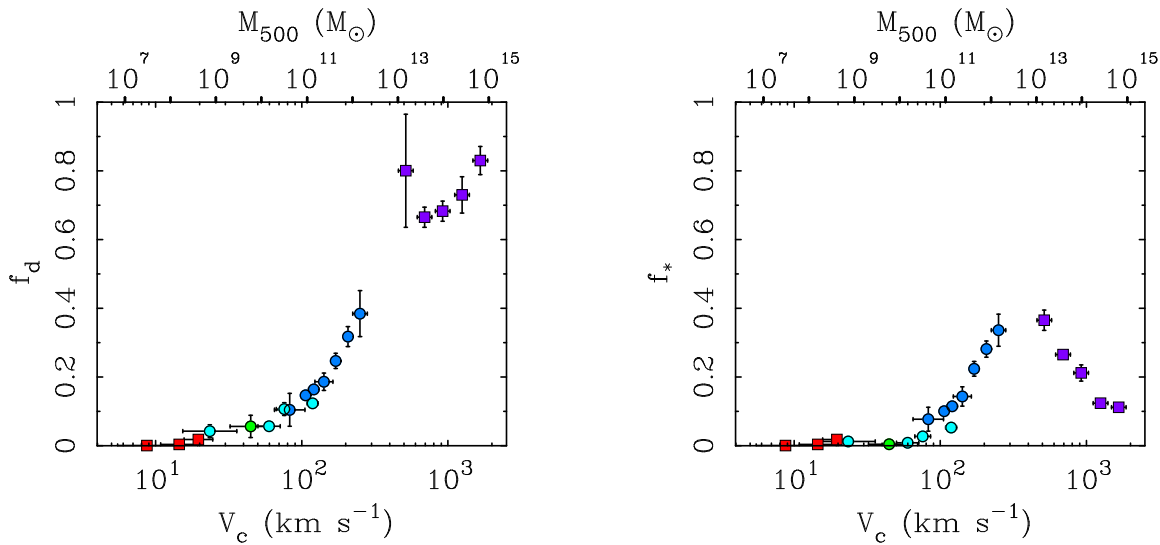}
\caption{The fraction of the expected baryons that are detected [left: ${f_d = M_b/(f_b M_{500})}$] 
and the fraction converted to stars [right: ${f_* = M_*/(f_b M_{500})}$].
Symbols have the same meaning as in Fig.~\ref{MbV} but with the binning of Table~\ref{bindat}.
The detected baryon fraction increases monotonically with mass while the stellar fraction peaks
between ${M}_{500} = 10^{12}$ and $10^{13}\;\mathrm{M}_{\sun}$.
\label{fdfstV}}
\end{figure*}

The choice of the radius $R_{500}$ is driven by the data, and is inevitably somewhat arbitrary.
A preferable choice might be the virial radius, which occurs around $\Delta \approx 100$ in \LCDM\ 
\citep{brynorm}.  The data do not constrain this, nor is there anything genuinely special about the
virial radius as the halo profile is expected to merge smoothly with the cosmic background.  So we
reference our masses to $R_{500}$, and caution that other plausible choices would give apparently
conflicting results merely because of the differing convention.  We also note that cluster
baryon fractions tend to be rising at the last measured point, which might help to reconcile them
with the cosmic baryon fraction.  In contrast, choosing a larger (e.g., virial) reference radius for
individual galaxies would only include more dark mass without any additional baryons, driving 
their inferred baryon fractions to smaller values.
\vfil
\subsection{Generalized Baryonic Tully-Fisher Relation}

Fig.~\ref{MbV} represents a generalization of the Baryonic Tully-Fisher relation to pressure
supported systems, and extension to higher and lower mass.
The data can be described by a broken power law, with a different slope and normalization
at the dwarf, disk galaxy, and cluster scale (Table~\ref{broken}). The variation between these
regimes is a small effect compared to the deviation from a constant baryon fraction 
(solid line in Fig.~\ref{MbV}).  The gravitating mass scales as
$M \sim V^3$, but the observed relation is steeper at all scales.
Consequently, the detected baryon mass
deviates systematically from the expectation of a constant cosmic baryon fraction.

\subsection{Binned Data}

The data of \citet{giodini} represent the binning of many individual clusters, while the remaining data
in Fig.~\ref{MbV} represent individual galaxies.  To place all data
on the same footing, we also bin the galaxy data.  
We bin in intervals in baryonic mass containing roughly equal numbers of galaxies.  We maintain
the distinction between galaxies of different types, and for the gas dominated galaxies, we maintain
separation between the data of \citet{stark} and \citet{trach}.  The binned data are reported in
Table~\ref{bindat} and plotted in Fig.~\ref{fdfstV}.  Table~\ref{bindat} gives the mean and standard
deviation in the mean of each logarithmic 
bin in circular velocity, baryonic mass, stellar mass, and gravitating mass
within $\Delta = 500$. It also gives the detected baryon fraction $f_d = M_b/(f_b M_{500})$, which is
the ratio of known baryons to the amount expected from the cosmic baryon fraction.  The fraction of
these nominally available baryons that have been converted into stars is given by the stellar fraction
$f_* = M_*/(f_b M_{500})$.  
\vfil
\section{Discussion}

The stellar fraction reaches a maximum between
${M}_{500} = 10^{12}$ and $10^{13}\;\mathrm{M}_{\sun}$ (Fig.~\ref{fdfstV}).
This is broadly consistent with previous results based on counting statistics \citep{yang}.
However, there may be some offset in the peak mass scale relating to the long standing
dichotomy between Tully-Fisher and luminosity function based normalizations of the halo
mass function.  Notably, the efficiency of star formation appears to increase monotonically
with mass for individual galaxies \citep{baldry}.  After the transition to cluster halos containing
many galaxies, the efficiency declines again.

Perhaps the most striking aspect of Fig.~\ref{fdfstV} is that the fraction of detected baryons
falls short of the cosmic fraction at all scales.  In no system is it
unity, as would be expected if we had a complete accounting of all the baryons associated with
a given bound system.  Where are all these missing baryons?

An obvious possibility is that the missing baryons are present, and simply inhabit their dark matter halos
in some undetected form.  Indeed, one might expect that not all baryons would have time to cool into the
observed cold gas and stellar component of galaxies.  In this case, many baryons might remain mixed
in with the dark halo.  

The notion that we are simply not seeing many or even most of the baryons in individual galaxies is
profoundly unsatisfactory.  Direct searches for hot halo gas have turned up nothing substantial:
halo baryon reservoirs fail to explain the observed deficit by two orders of magnitude \citep{bregman,andbreg}.  
In clusters, the hot gas is detected, and constitutes the majority of the baryons.
Clusters fall short of the cosmic baryon fraction by a modest amount \citep{mccarthy,giodini} which might
be readily explicable \citep{crain}.  However, the fraction of missing baryons is highly significant
on the scales of individual galaxies.
In dwarfs, fewer than 1\% of the baryons expected from the cosmic fraction are detected.

The detected baryon fraction varies systematically with scale.
It is a matter of taste whether one chooses to describe this scale as one of circular velocity,
mass, or potential well depth.  It is tempting to attribute this correlation to feedback processes
being more effective in objects with smaller potential wells \citep{DS}.  However, the details
are heinously complicated \citep{MayerMoore,crazy} and not well understood.
We would naively expect feedback to be a messy process resulting in lots of scatter in any
correlations that might result.  Instead, the observed relation is remarkably tight.  
In principle, the entire range $0 \le f_d \le 1$ is accessible at each mass, yet only a 
very particular value is observed.  Moreover, the current potential well is
the result of the hierarchical assembly of many smaller building blocks.  Left to itself,
a small dark matter halo will have a small $f_d$.  If incorporated into a larger halo,
$f_d$ goes up.  How does each building block know to bring the right amount of baryons 
to the final halo? 

We do not see a satisfactory solution to this missing baryon problem at present.
Considerable work remains to be done to obtain a complete understanding of the universe and its contents.

\acknowledgements  The work of S.S.M.\ is supported in part by NSF grant AST 0505956. 
 M.J.Z.\ was supported by an REU supplement from the NSF.
 The work of W.J.G.B.\ is based upon research supported by the South African
Research Chairs Initiative of the Department of Science and Technology and
National Research Foundation.

\end{document}